\documentclass[12pt,epsfig]{article}
\def\be{\begin{equation}}
\def\ee{\end{equation}}
\def\bear{\begin{eqnarray}}
\def\eear{\end{eqnarray}}
\def\bearst{\begin{eqnarray*}}
\def\eearst{\end{eqnarray*}}

\begin{document}
\begin{center}
{\large\bf  Area Quantization in Quasi-Extreme Black Holes}
\end{center}
\vspace{1ex}
\centerline{\large
E. Abdalla$^{\ast (1)}$, K. H. C. Castello-Branco$^{\ast (2)}$ and
A. Lima-Santos$^{\dag (3)}$}
\begin{center}
$^{\ast}$Universidade de S\~ao Paulo, Instituto de F\'\i sica\\
Caixa Postal 66318, 05315-970, S\~ao Paulo-SP, Brazil;\\
$^{\dag}$Universidade Federal de S\~ao Carlos, Departamento de
F\'{\i}sica \\
Caixa Postal 676, 13569-905, S\~ao Carlos-SP, Brazil\\
\end{center}
\vspace{6ex}

\begin{abstract}

We consider quasi-extreme Kerr and quasi-extreme Schwarzschild-de
Sitter black holes. From the known analytical expressions obtained for
their quasi-normal modes frequencies, we suggest an area quantization
prescription for those objects.

\end{abstract}
\vspace{4ex}
(1)eabdalla@fma.if.usp.br\\
(2)karlucio@fma.if.usp.br\\
(3)dals@df.ufscar.br
\newpage

The question of the quantization of the black hole horizon area is
well posed and
has been considered long ago by Bekenstein \cite{bekensteinlnc},
being a major issue since then \cite{bekenstmukanov}, \cite{kastrup},
\cite{bekenstescola}, \cite{hod}, \cite{dreyer}, \cite{kunsteter}, \cite{motl}, \cite{corichi}. The microscopic origin
of the black hole entropy \cite{bekensold}, \cite{hawking} is also an
unanswered question. There
are attempts to partially understand these questions using string theory
\cite{stromingervafa}, \cite{mazur} as well as the canonical approach
of quantum gravity \cite{thiemann}, \cite{ashtekarcorichi},
\cite{rovellismolin}, \cite{ashtekarlewandowski}. Recently, the
quantization of the black hole area has been considered \cite{hod},
\cite{dreyer} as a result of the absorption of
a quasi-normal mode excitation. Bekenstein's idea for
quantizing a black hole is based on the fact that its horizon area, in the
nonextreme case, behaves as a classical adiabatic invariant
\cite{bekensteinlnc}, \cite{bekenstescola}. It is worthwhile studying
how quasi-extreme holes would be quantized. It is specially
interesting to investigate this case since we analytically know the
quasi-normal mode spectrum of some black holes of that kind, namely the
quasi-extreme Kerr \cite{extremekerr} and quasi-extreme
Schwarzschild-de Sitter ({\it i.e.}, near-Nariai) \cite{carlemos}
solutions. The quasi-normal modes of black holes are the
characteristic, ringing frequencies which result
from their perturbations \cite{chandra} and provide a unique
signature of these objects \cite{kokkotas}, possible to be observed in
gravitational waves. Besides, quasi-normal modes
have been used to obtain further information of the space-time
structure, as for example in \cite{horohubeny}, \cite{birmingham}, \cite{carlemos2}, \cite{abdlimawang}, and \cite{abdlimakarl}.

Furthermore, gravity in such extreme configurations are an excellent
laboratory for the understanding of quantum gravity, and
information about the quantum structure of space-time can be derived in
such contexts by means of general setups \cite{abdbinsu}.

The first case of interest to us where the black hole quasi-normal
mode spectrum is analytically known is the quasi-extreme Kerr black
hole. In this case, the specific angular momentum of the hole, $a$, is
very nearly its mass M ($a \approx M$). Detweiler \cite{detweiler} was
able to show that in such a case there is an infinity of quasi-normal
modes given by \cite{extremekerr}
\be
\omega_n M\approx
\frac{m}{2}-\frac{1}{4m}\exp[\frac{\xi-2n\pi}{2\delta}+i\eta]\,,
\label{omegakerrextremo}
\ee
where $n=0,1,...\,$ labels the solution, $m$ is an integer labeling
the axial mode of the perturbation, while $\xi, \delta\,$, and $\eta$
are constants. We note that (\ref{omegakerrextremo}) is valid for
$\ell = m\,$, where $\ell$ is the multipole index of the
perturbation. For details we refer the reader to \cite{extremekerr}.

In Boyer-Lindq\"uist coordinates the Kerr solution reads
\bear
&&ds^{2}= -(1-\frac{2Mr}{\Sigma})dt^{2} -
\frac{4Mar\sin^{2}\theta}{\Sigma}dtd\phi + \frac{\Sigma}{\Delta}dr^{2}
+ \Sigma d\theta^{2} + \nonumber \\
&& + \left(r^{2} + a^{2} + 2Ma^{2}r\sin^{2}\theta
\right)\sin^{2}\theta d\phi^{2},
\eear
where
\be
\Delta=r^{2} - 2Mr + a^{2} \,,
\ee
\be
\Sigma= r^{2} + a^{2}\cos^{2}\theta \,.
\ee
$M$ and $0\leq a \leq M$ are the black hole mass and specific angular
momentum ($a=J/M$), respectively. The horizons are at
$r_{\pm}= M \pm \sqrt{M^{2}-a^{2}}\,$.

In units $G=c=1$, the black hole horizon area and its surface gravity
(temperature) are given, respectively, by
\be
A= 4\pi (r_+^2 + a^2)\,.
\label{area1}
\ee
\be
\kappa = \frac{1}{4A}(r_+ - r_-)\,.      \label{temp}
\ee

Based on {\it Bohr's correspondence principle} (``for large quantum
numbers, transition frequencies should equal classical frequencies''), Hod
\cite{hod} has considered the asymptotic limit $n \rightarrow \infty$
for the quasi-normal mode frequencies $\omega_n$ of a Schwarzschild
black hole in order to determine the spacing of its equally spaced
quantum area spectrum. That asymptotic quasi-normal mode spectrum was
obtained numerically by Nollert \cite{norlett}. Recently, Motl
\cite{motl} has computed analytically $Re(\omega_n)$ as
$n \rightarrow \infty$, finding agreement with the numerical value of
Nollert. In this large $n$ limit, Hod \cite{hod} then assumed that a
Schwarzschild black hole mass should increase by $\delta M=\hbar
Re(\omega_n)$ when it absorbs a quantum of energy $\hbar Re(\omega_n)$.

As in Ref. \cite{hod}, we expect that the real part of the quasi-normal
mode frequency for large $n$ corresponds to an addition of energy equal to
$\hbar Re(\omega_n)$ to the quasi-extreme Kerr black hole mass as it
falls into its event horizon. Then, taking the limit $n \rightarrow
\infty$ for $\omega_n$ in (\ref{omegakerrextremo}) we simply have
\be
\omega_n \approx \frac{m}{2M} \quad, \quad n \rightarrow \infty\,.
\ee
Contrary to the Schwarzschild case, where the limit
$n \rightarrow \infty$ gives highly damped modes, for the present case,
it gives virtually {\it undamped} modes with frequencies close to
the upper limit of the superradiance interval \cite{super-rad}, $0 < \omega <
m\Omega\,$, where $\Omega= 4\pi a/A$ is angular velocity of the
horizon. The quasi-normal mode spectrum (\ref{omegakerrextremo}) of
near-extreme Kerr black holes leads to interesting
consequences, as recently analysed by Glampedakis and Anderson
\cite{extremekerr}.

Furthermore, here the angular momentum $\hbar m$ adds to the angular
momentum $J=Ma$ associated with the Kerr solution. We then have a pair
of variations for black hole parameters given by
\be
\delta M = \frac{\hbar m}{2M}\quad; \qquad \delta J =\hbar m \qquad \Rightarrow
\qquad \delta a =\hbar m(\frac{1}{M}-\frac{a}{2M^2})\,.
\label{variacoes}
\ee
In what follows we will consider $\hbar=1$ and for the sake of brevity
$m=1$.

The variation of the horizon area is related to the first law of black hole
thermodynamics,
\be
\delta M = \kappa \delta A + \Omega \delta J \,.   \label{1alei}
\ee

Making use of relations (\ref{variacoes}), we can obtain from
(\ref{area1}) that, for a near-extreme hole ($a \approx M\,$), the area
variation is given by
\be
\delta A = 8\pi \left (1+ \sqrt{\frac{M-a}{2M}}\right)\quad,
\label{varar1}
\ee
up to first order in $(M-a)^{1/2}$.

Therefore, for strictly extreme holes, we simply have $\delta A =
8\pi\,$.

For $a \approx M\,$, we can express $\kappa$ and $\Omega$, respectively, as
\be
\kappa \approx \frac{1}{16\pi} \frac{\sqrt{M^2-a^2}}{M^2}\left
[1-\frac{\sqrt{M^2-a^2}}{M} \right ]
\label{vartemp}
\ee
and
\be
\Omega \approx \frac{a}{2M^2}\left (1-\frac{\sqrt{M^2-a^2}}{M}+
\frac{M^2-a^2}{M^2} \right)\,.
\label{varveloang}
\ee

Finally, from (\ref{variacoes}), (\ref{varar1}), (\ref{vartemp}),
and (\ref{varveloang}), to order $(M-a)^{3/2}$, we obtain
\be
\kappa \delta A + \Omega \delta J \approx \frac{1}{2M}\,, \label{concorda}
\ee
in agreement with the first law of black hole thermodynamics
(\ref{1alei}). Thus we can prescribe the quantization of
a quasi-extreme Kerr black hole area as
\be
A_n = n \delta A\ell^2_P\simeq 8\pi\ell^2_P\,n\,,
\label{arquant1}
\ee
where $n=1,2,...$ and $\ell_P$ is the Planck length.

A second case where we can obtain information about the black hole
parameters involved in its quantization is the near-extreme
Schwarzschild-de Sitter (S-dS) black hole. This is the case when the
mass of the black hole is increased as to arrive near the limit
$M_N=R/3\sqrt{3}\,$, where the constant $R$ is related to the
cosmological constant $\Lambda$ by $R^2=3/\Lambda$. This is the Nariai limit
\cite{nariai}, for which
the black hole and cosmological horizons coincide. The S-dS metric is
\cite{gibhawk}
\be
ds^2= -f(r)dt^2 + f(r)^{-1}dr^2 + r^2(d\theta^2+\sin^2\theta d\phi^2)\,,
\ee
where
\be
f(r)= 1-\frac{2M}{r}-\frac{r^2}{R^2}\,,
\ee
and $0\leq M \leq M_N$ is the black hole mass. The roots of $f(r)$ are
$r_b,r_c$ and $r_0 =-(r_b + r_c)$, where $r_b$ and $r_c$ are the black
hole and cosmological horizon radii, respectively.
To each horizon there is a surface gravity, given by $\kappa_{b,c}
=\frac{1}{2}\frac{df}{dr}|_{r=r_{b,c}}$. For $\kappa_b$ we have the expression
\be
\kappa_b = \frac{(r_c-r_b)(r_b-r_0)}{2R^2r_b}\,.   \label{surfgrav}
\ee

As in the Kerr case, we will perform the variation of $M\,$. It
is useful to write $M$ and $R$ in terms of $r_b$ and $r_c$ as
\be
2MR^2= r_br_c(r_b + r_c)\,,
\label{massa}
\ee
\be
R^2= r_b^2 + r_b r_c + r_c^2\,.
\label{raio}
\ee

The analytical quasi-normal mode spectrum for the
quasi-extreme S-dS black hole has been recently derived
by Cardoso and Lemos \cite{carlemos} and reads
\be
\omega_n =\kappa_b \left [\sqrt{\frac{V_0}{\kappa_b^2}-\frac{1}{4}}
-i(n+\frac{1}{2}) \right ] \,,
\label{modos2}
\ee
where $n=0,1,...\,$, and $V_0=\kappa_b^2\ell(\ell+1)$, for scalar and
electromagnetic perturbations, and $V_0=\kappa_b^2(\ell+2)(\ell-1)$ for
gravitational perturbations.

Since we are considering the near extreme limit of the S-dS solution,
for which $(r_c - r_b)/r_b << 1$,
it is suitable for our purposes to write the black hole mass as
\be
M = M_N + \mu = \frac{R}{3\sqrt{3}} + \mu \, .
\ee

Therefore, since $R=\sqrt{3/\Lambda}$ is fixed, the use of (\ref{massa})
and (\ref{raio}) leads us to
\be
\delta M = \delta \mu = \frac{r_b \Delta r \delta r_b}{2R^2}\,,
\label{varmass1}
\ee
where $\Delta r= r_c - r_b\,$.

Similarly as we did for the Kerr case, here we can consider
$\delta M=\hbar Re(\omega_n)\,$ and in view of (\ref{modos2}) and
(\ref{surfgrav}) write ($\hbar=1$)
\be
\delta M= \frac{\Delta r}{2r_b{^2}}\,\sqrt{(\ell+2)(\ell-1)-\frac{1}{4}} \quad,
\label{varmass2}
\ee
where we have used $R^2\sim 3r_b^2 $ and considered $V_0$ for gravitational perturbations.

The variation of the black hole horizon area,
\be
\delta A_b = 8\pi r_b\delta r_b \,,
\ee
then gives us
\be
\delta A_b = 24\pi\sqrt{(\ell+2)(\ell-1)-\frac{1}{4}}\quad,
\label{varar2}
\ee
for gravitational quasi-normal modes and $\delta A_b =
24\pi\sqrt{\ell(\ell+1)-\frac{1}{4}}$ otherwise.

Thus we can prescribe the quantum area spectrum for a quasi-Nariai
black hole as
\be
A_{b_{n}} = n \delta A{_b} \ell^2_P \simeq 12\pi \sqrt{15}\ell^2_P\, n\,,
\label{arquant2}
\ee
where $n=1,2,...\,$, or, in the case of scalar or electromagnetic
perturbations, $12\pi \sqrt{7}\ell^2_P\,n\,$.

In summary, with the knowledge of the analytical quasi-normal mode
spectrum of near extreme Kerr and near extreme S-dS black
holes, as given in \cite{extremekerr} and \cite{carlemos}, we have
prescribed how their horizon area would be quantized. This was done by
simply assuming they have a uniformly spaced area spectrum given by
$A_n = \delta A \ell^2_P\, n\,$, where $\delta A$ is the area variation
caused by absorption of a quasi-normal mode. This was done in analogy
with the Schwarzschild case, where the spacing of its area spectrum
was determined by means of the knowledge of its asymptotic (``large $n$
``) quasi-normal mode frequencies \cite{hod}. In the cases regarded
here, the results for the spacing of the area spectrum differ from
that for Schwarzschild, as well as for non-extreme Kerr \cite{hod2}
black holes, in which cases, the spacing is predicted to be
given by $4ln\,3$. This factor comes from the real part of the
asymptotic quasi-normal mode frequencies of those black holes
\cite{hod}, \cite{hod2}. Such a difference may be justified due to the
quite different nature of the asymptotic quasi-normal mode spectrum of
the near extreme black holes we considered. Furthermore, it should be
no {\it a priori} reason for expecting the same behaviour for the
asymptotic quasi-normal mode frequencies of near extreme and
non-extreme black holes.


\bigskip
\noindent {\bf Acknowledgements}\par
The authors would like to thank {\bf CNPq-Brazil} ({\it Conselho
Nacional de Desenvolvimento Cient\'\i fico e Tecnol\'ogico}) and
{\bf FAPESP-Brazil} ({\it Funda\c{c}\~ao de Amparo \`a Pesquisa
no Estado de S\~ao Paulo}) for financial support.


\end{document}